\documentclass[%
 reprint,
 amsmath,amssymb,
 aps,
 superscriptaddress,
pra,
]{revtex4-1}

\usepackage{graphicx}
\usepackage{dcolumn}
\usepackage{bm}

\begin{document}

\preprint{APS/123-QED}

\title{Interferences between Bogoliubov excitations and their impact on the evidence of superfluidity in a paraxial fluid of light}

\author{Quentin Fontaine}
\affiliation{Laboratoire Kastler Brossel, Sorbonne Universit\'e, CNRS, ENS-PSL Research University, Coll\`ege de France, Paris 75005, France}
\author{Pierre-\'Elie Larr\'e}
\affiliation{CY Cergy Paris Universit\'e, CNRS, LPTM, F-95000 Cergy, France}
\author{Giovanni Lerario}
\affiliation{Laboratoire Kastler Brossel, Sorbonne Universit\'e, CNRS, ENS-PSL Research University, Coll\`ege de France, Paris 75005, France}
\affiliation{CNR NANOTEC Institute of Nanotechnology, Via Monteroni, 73100 Lecce, Italy}
\author{Tom Bienaim\'e}
\affiliation{Laboratoire Kastler Brossel, Sorbonne Universit\'e, CNRS, ENS-PSL Research University, Coll\`ege de France, Paris 75005, France}
\author{Simon Pigeon}
\affiliation{Laboratoire Kastler Brossel, Sorbonne Universit\'e, CNRS, ENS-PSL Research University, Coll\`ege de France, Paris 75005, France}
\author{Daniele Faccio}
\affiliation{School of Physics and Astronomy, University of Glasgow, Glasgow G12 8QQ, UK}
\author{Iacopo Carusotto}
\affiliation{INO-CNR BEC Center and Universit\`a di Trento, Via Sommarive, 38123 Povo, Italy}
\author{\'Elisabeth Giacobino}
\affiliation{Laboratoire Kastler Brossel, Sorbonne Universit\'e, CNRS, ENS-PSL Research University, Coll\`ege de France, Paris 75005, France}
\author{Alberto Bramati}
\affiliation{Laboratoire Kastler Brossel, Sorbonne Universit\'e, CNRS, ENS-PSL Research University, Coll\`ege de France, Paris 75005, France}
\author{Quentin Glorieux}
\email[Corresponding author:] {quentin.glorieux@lkb.upmc.fr}
\affiliation{Laboratoire Kastler Brossel, Sorbonne Universit\'e, CNRS, ENS-PSL Research University, Coll\`ege de France, Paris 75005, France}


\date{\today}

\begin{abstract}
Paraxial fluids of light represent an alternative platform to atomic Bose-Einstein condensates and superfluid liquids for the study of the quantum behaviour of collective excitations.
A key step in this direction is the precise characterization of the Bogoliubov dispersion relation, as recently shown in two experiments \cite{Vocke2015, Fontaine2018}.
However, the predicted interferences between the phonon excitations that would be a clear signature of the collective superfluid behaviour  have not been observed to date.
Here, by analytically, numerically, and experimentally exploring the phonon phase-velocity, we observe the presence of interferences between counter-propagating Bogoliubov excitations and demonstrate their critical impact on the measurement of the dispersion relation.
These results are evidence of a key signature of light superfluidity  and provide a novel characterization tool for  quantum simulations with photons.
\end{abstract}

\maketitle

\section{Introduction}

The weakly beyond-mean-field description of a Bose quantum fluid, initially introduced by Bogoliubov, relies on small collective excitations on top of a time-independent condensate \cite{Bogoliubov1947, Pitaevskii2016}.
These excitations are described as non-interacting quasi-particles with a specific energy spectrum: sound-like at low momenta and free-particle-like at large momenta.
Due to these linear then parabolic dependences at respectively low and large momenta, a system exhibiting this type of dispersion  satisfies the Landau criterion for superfluidity \cite{leggett2001bose}, which is a benchmark for the system to behave as a superfluid, one of the most striking manifestations of quantum many-body physics. 

In optics, a growing community focuses on quantum-fluid physics with light in non-linear media \cite{carusotto2013quantum}.
For example, Bose-Einstein condensation has been observed both in  exciton-polariton \cite{kasprzak2006bose} and dye-filled \cite{klaers2010bose} microcavities.
Initially proposed by Pomeau and Rica \cite{Pomeau1993} and neglected experimentally for a long time, paraxial fluids of light present exciting perspectives for studying quantum-fluids physics \cite{carusotto2014superfluid,noh2016quantum}.
In this approach, photons acquire an effective mass as a consequence of the paraxial approximation while effective repulsive photon-photon interactions are mediated by the optical non-linearity of the medium in which they propagate.
Experimental implementations rely on the propagation of an intense laser beam within a negative third-order, Kerr non-linear medium such as photorefractive crystals \cite{wan2007dispersive,michel2018superfluid}, thermo-optic media \cite{vocke2016role, Elazar2013}, and hot atomic vapors \cite{vsantic2018nonequilibrium, Fontaine2018}.
In this 2D + 1 geometry, the system is two-dimensional in the transverse direction and the propagation coordinate is analogous to an effective time.

Recently, two experiments have measured the dispersion relation of weak-amplitude excitations on top of a paraxial fluid of light with two complementary approaches \cite{Vocke2015, Fontaine2018} following a proposal of Ref. \cite{carusotto2014superfluid}.
The evolution of these elementary excitations is described by the Bogoliubov theory, revealing the rich analogy existing between non-linear photonics and quantum condensed matter physics.
If this analogy is now well established,  theoretical works \cite{Larre2017,Ferreira2018} have questioned the presence and the impact of interferences between counter-propagating Bogoliubov excitations in paraxial fluids of light.
In this paper, we present the first experimental evidence of these interferences and we demonstrate their dramatic impact on the reconstruction of the dispersion relation and on the identification of superfluidity of light.
Moreover, we propose an interpretation of these interferences as stimulated analogue of the Sakharov oscillations of cosmology~\cite{sakharov1966initial,cosmobook}, recently observed in atomic condensate~\cite{hung2013cosmology}.

Finally, we show that this effect is robust across several experimental systems used for paraxial fluids of light by numerically taking into account the photon absorption, the finite size of the fluid, the saturation and the non-locality of the photon-photon interactions.
Because all these corrections only marginally impact the observed behavior, our work opens the way to novel experimental techniques for probing paraxial fluid of light, based on the observation of Bogoliubov-excitation interferences.
For example, we propose that extracting the contrast of constructive interference fringes in the output plane as a function of the probe parameters will give access to the efficiency at which we can excite phonons, also known as the static structure factor \cite{shammass2012phonon}.

\section{Paraxial fluid of light}
\label{TheoreticalGrounds}

We consider a monochromatic beam of light propagating  along the positive-$z$ direction in a $\smash{\chi^{(3)}}$ non-linear medium.
In the paraxial and scalar approximations, the evolution of the slowly varying envelope $\mathcal{E}(\mathbf{r}_{\perp},z)$ of the complex electric field $E(\mathbf{r}_{\perp},z)=\mathcal{E}(\mathbf{r}_{\perp},z)e^{i(k_0z-\omega t)}$ is known to obey the non-linear Schr\"{o}dinger equation (NLSE) of non-linear optics \cite{Boyd2008}:
\begin{equation}
    i\frac{\partial\mathcal{E}}{\partial z}=\bigg({-}\frac{1}{2k_0}\boldsymbol{\nabla}_{\perp}^{2}-\frac{3k_0\chi^{(3)}}{8n^{2}}|\mathcal{E}|^{2}-\frac{i\alpha}{2}\bigg)\mathcal{E}.
    \label{NLSE}
\end{equation}
In this equation, $k_0=n\omega/c$ is the laser propagation constant in the medium with $n$ the linear refractive index, $\omega$ the laser angular frequency, and $c$ the vacuum speed of light, $\boldsymbol{\nabla}_{\perp}$ is the gradient with respect to the transverse coordinates $\mathbf{r}_{\perp}=(x,y)$, and $\alpha\geqslant0$ is the absorption coefficient describing photon losses.

\begin{figure}[t!]
\centering
\includegraphics[width=6.5cm]{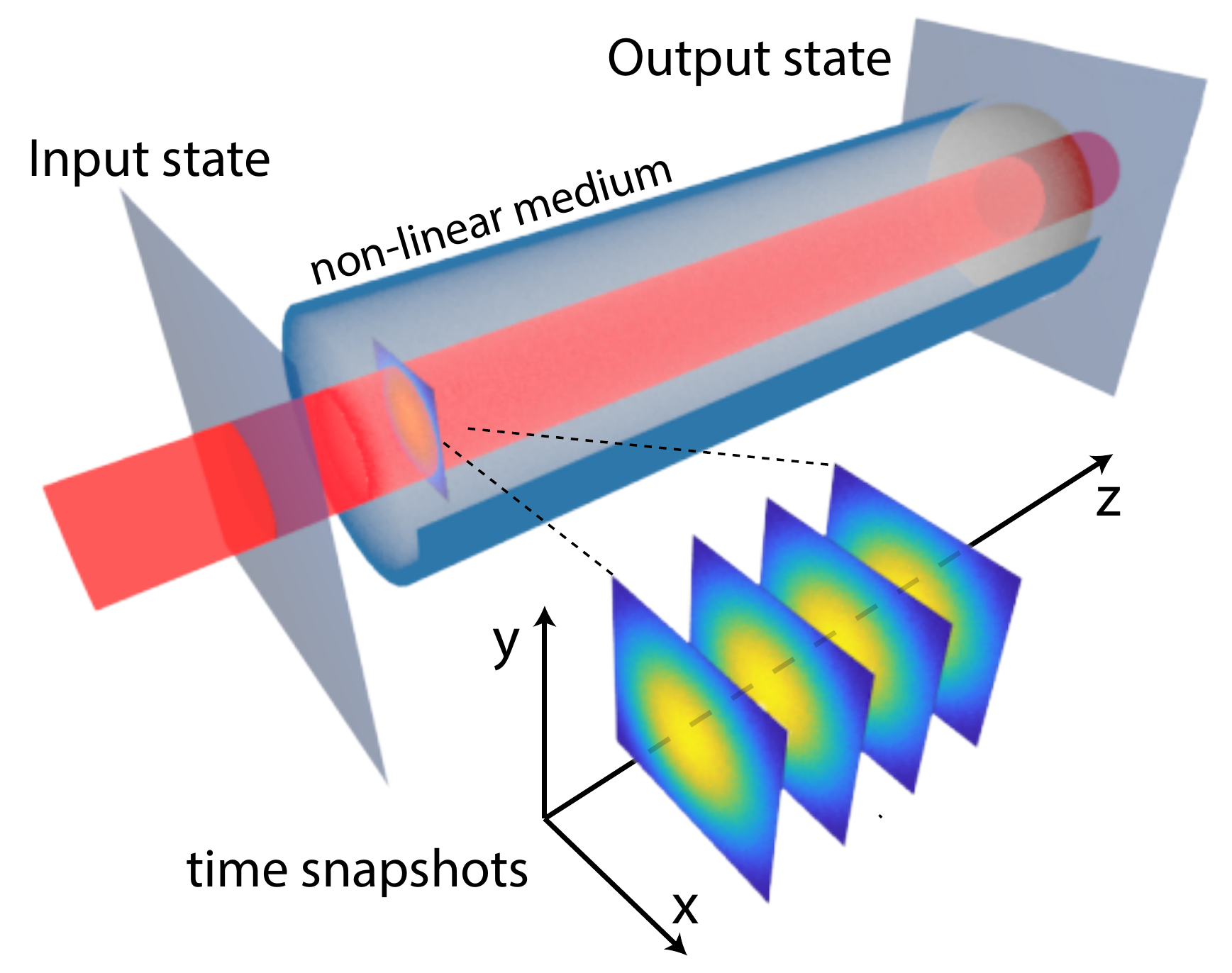} 
\caption{Paraxial fluid of light.
In the paraxial and scalar approximations, a laser beam propagates along the $z$ axis in a $\smash{\chi^{(3)}}$ non-linear medium according to the effective Gross-Pitaevskii equation (\ref{NLSE}).
The field profile on each transverse $\mathbf{r}_{\perp}=(x,y)$ plane along the propagation direction $z$ is equivalent to a snapshot of the evolution of the paraxial fluid of light.}
\label{fig:manip}
\end{figure}

Except for the last term describing losses, \eqref{NLSE} is formally analogous to the Gross-Pitaevskii equation (GPE) describing the temporal evolution of the macroscopic wavefunction of an atomic Bose-Einstein condensate (BEC) in two dimensions \cite{Pitaevskii2016}.
In the right-hand side in particular, the Laplacian term mimics the kinetic-energy term with a mass corresponding to the laser propagation constant $k_0$.
In addition, the $\smash{\chi^{(3)}}$ contribution is analogous to the contact-interaction potential with an interaction parameter $g$ proportional to the Kerr susceptibility:
$ g=-3k_0\chi^{(3)}/(8n^{2})$.
In the following, we consider the non-linearity to be self-defocusing ($\smash{\chi^{(3)}}<0$) so that the effective photon-photon interactions are repulsive ($g>0$).
In this analogy, the fluid density $\rho(\mathbf{r}_{\perp},z)$ is directly proportional to the field intensity $I(\mathbf{r}_{\perp},z)$ according to
$
\rho(\mathbf{r}_{\perp},z)=|\mathcal{E}(\mathbf{r}_{\perp},z)|^{2}=2I(\mathbf{r}_{\perp},z)/(c\epsilon_{0}n) 
$,
 where $\epsilon_{0}$ denotes the vacuum permittivity. However, while the GPE describes the evolution of a condensate wavefunction for a matter quantum fluid \textit{in time}, the NLSE describes how the electric-field envelope $\mathcal{E}(\mathbf{r}_{\perp},z)$ of the light beam propagates \textit{in space}, along the $z$ axis.
Therefore, the propagation coordinate $z$ is equivalent to an effective time in the NLSE.
As a consequence, every transverse plane (spanned by $\mathbf{r}_{\perp}$) along the propagation axis $z$ can be regarded as a snapshot of the evolution of the two-dimensional paraxial fluid of light (see Fig.~\ref{fig:manip}). 
The role of the physical time $t$ as a third spatial coordinate for propagating light was highlighted in~\cite{Larre2015}. 
These features are however not relevant in the monochromatic excitation case under investigation here.

In the following theoretical description of the paraxial fluid of light and of its Bogoliubov excitations, we disregard the effect of photon losses by taking $\alpha=0$.
This approach has the advantage  of shining light on the general features without harming the generality of our conclusions. A complete theory including photon losses will be presented later in Fig.~\ref{fig:ShiftVersus}, showing no qualitative change.

In the ideal lossless case, we assume that the beam maintains a wide flat-top and $z$-independent intensity profile all along its propagation, so that the corresponding solution of \eqref{NLSE} reads 
\begin{equation}
    \label{FlatTopBeam}
    \mathcal{E}_{0}(z)=\sqrt{\rho_{0}}e^{-ik_0\Delta nz}.
\end{equation}
In this equation, $\rho_{0}$ is the density of the homogeneous paraxial fluid of light and $\Delta n=g\rho_{0}/k_0$ is the change of refractive index induced by the optical non-linearity.
A small departure from the uniform and stationary configuration (\ref{FlatTopBeam}) is described by a solution of \eqref{NLSE} of the form
\begin{equation}
    \mathcal{E}(\mathbf{r}_{\perp},z)=\mathcal{E}_{0}(z)+\delta\mathcal{E}(\mathbf{r_{\perp}},z),
\end{equation}
where $|\delta\mathcal{E}(\mathbf{r_{\perp}},z)|\ll|\mathcal{E}_{0}(z)|$.
Such an expansion depicts weak-amplitude fluctuations (for example intensity fluctuations) on top of the homogeneous (i.e., $\mathbf{r}_{\perp}$-independent) background defined in \eqref{FlatTopBeam}. 
The complex field $\delta\mathcal{E}(\mathbf{r}_{\perp},z)$, which is solution of the linearized version of \eqref{NLSE}, can be decomposed following the Bogoliubov approach \cite{Pitaevskii2016} as a linear superposition of plane waves counter-propagating in the transverse $\mathbf{r}_{\perp}$ plane with opposite wavevectors $\pm\mathbf{k}_{\perp}$ and oscillating in the effective time $z$ at the same angular frequency $\Omega_{\mathrm{B}}(\mathbf{k}_{\perp})$:
\begin{align}
    \notag
    \delta\mathcal{E}(\mathbf{r}_{\perp},z)&\left.=e^{-ik_0\Delta nz}  \int\frac{d^{2}\mathbf{k}_{\perp}}{(2\pi)^{2}}\Big\{u(\mathbf{k}_{\perp})b_{\mathbf{k}_{\perp}}e^{i[\mathbf{k}_{\perp}\cdot\mathbf{r}_{\perp}-\Omega_{\mathrm{B}}(\mathbf{k}_{\perp}) z]}\right. \\
    \label{Fluctuation}
    &\left.\hphantom{=}+v^{\ast}(\mathbf{k}_{\perp})b^{\ast}_{\mathbf{k}_{\perp}}e^{-i[\mathbf{k}_{\perp}\cdot\mathbf{r}_{\perp}-\Omega_{\mathrm{B}}(\mathbf{k}_{\perp}) z]}\Big\}.\right.
\end{align}
In this expression, the complex amplitudes of the plane waves with wavevectors $\mathbf{k}_{\perp}$ and $-\mathbf{k}_{\perp}$ are respectively denoted by $u(\mathbf{k}_{\perp})b_{\mathbf{k}_{\perp}}$ and $v^{\ast}(\mathbf{k}_{\perp})b_{\mathbf{k}_{\perp}}^{\ast}$, where $b_{\mathbf{k}_{\perp}}$ is chosen to be homogeneous to a voltage times a length so that $u(\mathbf{k}_{\perp})$ and $v(\mathbf{k}_{\perp})$ are by construction dimensionless.
The latter satisfy the eigenvalue problem \cite{Pitaevskii2016}
\begin{equation}
    \label{BdG}
    \begin{split}
        &\mathcal{L}(\mathbf{k}_{\perp})
        \renewcommand\arraystretch{1}
        \begin{bmatrix}
            u(\mathbf{k}_{\perp}) \\
            v(\mathbf{k}_{\perp})
        \end{bmatrix}
        =\Omega_{\rm B}(\mathbf{k}_{\perp})
        \renewcommand\arraystretch{1}
        \begin{bmatrix}
            u(\mathbf{k}_{\perp}) \\
            v(\mathbf{k}_{\perp})
        \end{bmatrix}
        ,\quad\text{where} \\
        &\mathcal{L}(\mathbf{k}_{\perp})=
        \renewcommand\arraystretch{1}
        \begin{bmatrix}
            k_{\perp}^{2}/(2k_{0})+k_{0}\Delta n & \!\!\!
            k_{0}\Delta n \\
            -k_{0}\Delta n & \!\!\!
            -k_{\perp}^{2}/(2k_{0})-k_{0}\Delta n
        \end{bmatrix},
    \end{split}
\end{equation}
with the wavenumber $k_{\perp}=|\mathbf{k}_{\perp}|$.
Without loss of generality, we  take $u(\mathbf{k}_{\perp})$ and $v(\mathbf{k}_{\perp})$ to be real.
Setting the normalization condition $u^{2}(\mathbf{k}_{\perp})-v^{2}(\mathbf{k}_{\perp})=1$, we get the dispersion relation
\begin{align}
    \label{DispersionRelation}
    \Omega_{\mathrm{B}}(\mathbf{k}_{\perp})&=\sqrt{\frac{k_{\perp}^{2}}{2k_0}\bigg(\frac{k_{\perp}^{2}}{2k_0}+2k_0\Delta n\bigg)}\quad\text{and} \\
    \label{BogAmplitudes}
    u(\mathbf{k}_{\perp})\pm v(\mathbf{k}_{\perp})&=\bigg[\frac{k_{\perp}^{2}}{2k_0}\bigg/\Omega_{\mathrm{B}}(\mathbf{k}_{\perp})\bigg]^{\pm\frac12}.
\end{align}
Equation (\ref{DispersionRelation}) is the optical analog of the Bogoliubov excitation spectrum of an homogeneous two-dimensional atomic BEC at rest and  \eqref{BogAmplitudes} gives the $\mathbf{k}_{\perp}$ dependence of the Bogoliubov amplitudes $u(\mathbf{k}_{\perp})$ and $v(\mathbf{k}_{\perp})$. 
Here, the linear combinations $u+v$ and $u-v$ respectively correspond to the density and phase amplitudes of the Bogoliubov collective wave in wavevector space.


%
From \eqref{DispersionRelation}, we can extract the peculiar behavior of the Bogoliubov dispersion relation $\Omega_{\mathrm{B}}(\mathbf{k}_{\perp})$, which is linear (sound-like) at small $\mathbf{k}_{\perp}$ and parabolic (free-particle-like) at large $\mathbf{k}_{\perp}$:
\begin{equation}
    \Omega_{\mathrm{B}}(\mathbf{k}_{\perp})\simeq
    \begin{cases}
    c_{\mathrm{s}}k_{\perp} & \text{when} \quad k_{\perp}\xi\ll1 \\
    \displaystyle{\frac{k_{\perp}^{2}}{2k_0} + k_0 \Delta n} & \text{when} \quad k_{\perp}\xi\gg1.
    \end{cases}
\end{equation}
These asymptotic behaviors bring up the optical analogs of the Bogoliubov sound velocity, $c_{\mathrm{s}}=\sqrt{\Delta n}$, and of the healing length, $\xi=1/(k_0\sqrt{\Delta n})=1/(k_0c_{\mathrm{s}})$, of atomic BECs.
In the present optical context, $c_{\mathrm{s}}$ is by  construction dimensionless, as it corresponds to the propagation angle with respect to the $z$ axis.
The peculiar refraction properties corresponding to the constant $c_s$ in the $k \xi \to 0$ limit were highlighted in~\cite{Fontaine2018}.
In the large $k\xi$ limit, the shift in $\Omega_B(\mathbf{k}_\perp)$ is simply linked to the nonlinear refractive index change.


\section{Extracting the Bogoliubov dispersion relation from the phase velocity}


The phase velocity $v_{\mathrm{ph}}(\mathbf{k}_{\perp})$ of a Bogoliubov plane wave with wavevector $\mathbf{k}_{\perp}$ is related to the Bogoliubov dispersion relation $\Omega_{\mathrm{B}}(\mathbf{k}_{\perp})$ through
\begin{equation}
    v_{\mathrm{ph}}(\mathbf{k}_{\perp})=\frac{\Omega_{\mathrm{B}}(\mathbf{k}_{\perp})}{k_{\perp}}.
\end{equation}
Therefore, it is expected that we can directly reconstruct $\Omega_{\mathrm{B}}(\mathbf{k}_{\perp})$ from the measurement of $v_{\mathrm{ph}}(\mathbf{k}_{\perp})$, which can be assessed from the measurement of the distance
\begin{equation}
    S(\mathbf{k}_{\perp})=v_{\mathrm{ph}}(\mathbf{k}_{\perp})L
\end{equation}
that the Bogoliubov excitation travels in the transverse plane between the effective times $z=0$ and $z=L$, where $L$ stands for the length of the non-linear medium.

In the experimental configuration initially proposed in Ref. \cite{Vocke2015} and studied here, $S(\mathbf{k}_{\perp})$ corresponds to the transverse displacement of a weak interference pattern obtained by overlapping a large-intensity flat-top background with a low-intensity probe, slightly tilted by an angle $\theta_{\mathrm{i}}$ with the $z$ axis along which the background propagates.
These two beams come from the same laser, have the same frequency and polarization, and thus interfere, producing a small fluctuation $\delta\mathcal{E}(\mathbf{r}_{\perp},z)$ on top of the background envelope $\mathcal{E}_{0}(z)$ in the non-linear medium.
The norm $k_{\perp}=(k_0/n)\sin\theta_{\mathrm{i}}$ of the transverse wavevector of the incident probe is controlled by changing $\theta_{\mathrm{i}}$, which must be small enough so that the whole optical system falls into the paraxial limit $k_{\perp}\ll k_0$ considered here.

After propagation inside the medium of length $L$, the background (``bg'') and the probe (``p'') have accumulated different phases $\Phi_{\mathrm{bg}}$ and $\Phi_{\mathrm{p}}(\mathbf{k}_{\perp})$. According to \eqref{Fluctuation}, the latter depends on the Bogoliubov dispersion relation $\Omega_{\mathrm{B}}(\mathbf{k}_{\perp})$. The difference
\begin{equation}
    \Phi(\mathbf{k}_{\perp})=\Phi_{\mathrm{p}}(\mathbf{k}_{\perp})-\Phi_{\mathrm{bg}}
    \label{eqshift}
\end{equation}
between these two phases is responsible for an interference pattern in the transverse plane, shifted by
\begin{equation}
    \label{S-Phi}
    S(\mathbf{k}_{\perp})=\frac{\Phi(\mathbf{k}_{\perp})}{k_{\perp}}.
\end{equation}
Experimentally, it is possible to have access to
\begin{equation}
    \label{DeltaS}
    \Delta S(\mathbf{k}_{\perp})=S_{\mathrm{NL}}(\mathbf{k}_{\perp})-S_{\mathrm{L}}(\mathbf{k}_{\perp}),
\end{equation}
the relative deviation between the fringes patterns obtained at high and low background intensity, that is, in the non-linear (``NL'') regime and the linear (``L'') one, respectively.
This quantity can be, at first, estimated in a geometrical approach, as detailed below.

\subsection{Geometrical approach}

In the linear regime, simple geometry yields the following expressions for the phases accumulated by the background and the probe beams:
\begin{align}
    \Phi_{\mathrm{bg},\mathrm{L}}&=k_0L\quad\text{and} \\
    \Phi_{\mathrm{p},\mathrm{L}}(\mathbf{k}_{\perp})&=k_0\sqrt{L^{2}+L^{2}\tan^{2}\theta_{\mathrm{r}}}\simeq k_0L\bigg(1+\frac{\theta_{\mathrm{r}}^{2}}{2}\bigg),
\end{align}
where $\theta_{\mathrm{r}}\simeq\theta_{i}/n$ is the refraction angle of the probe at the entrance of the medium. Using $k_{\perp}\simeq(k_0/n)\theta_{\mathrm{i}}$, we then obtain 
\begin{equation}
    \label{LPPhaseDiff}
    \Phi_{\mathrm{L}}(\mathbf{k}_{\perp})=\frac{k_{\perp}^{2}}{2k_0}L.
\end{equation}
In a geometrical approach, the same formula is supposed to hold in the non-linear regime provided the free-particle dispersion relation $k_{\perp}^{2}/(2k_0)$ is replaced with the Bogoliubov spectrum (\ref{DispersionRelation}) and we obtain using \eqref{eqshift}:
\begin{equation}
    \Phi_{\mathrm{NL}}(\mathbf{k}_{\perp})=\Omega_{\mathrm{B}}(\mathbf{k}_{\perp})L.
    \label{HPPhaseDiff}
\end{equation}
In the light of Eqs.~(\ref{S-Phi}) and (\ref{DeltaS}), this geometric approach then leads to
\begin{equation}
    \Delta S(\mathbf{k}_{\perp})=\frac{k_{\perp}}{2k_0}\left[\sqrt{1+\Delta n\bigg(\frac{2k_0}{k_{\perp}}\bigg)^{2}}-1\right]L.
    \label{ShiftFaccio}
\end{equation}
This expression states that $\Delta S(\mathbf{k}_{\perp})$ saturates to a constant value proportional to the Bogoliubov speed of sound in the deep phonon regime:
\begin{equation}
    \label{SaturationFaccio}
    \Delta S(\mathbf{k}_{\perp})\underset{k_{\perp}\xi\ll1}{\simeq}\sqrt{\Delta n}L=c_{\mathrm{s}}L.
\end{equation}
This approach has been experimentally implemented for non-local photon fluids \cite{Vocke2015}.
In particular \eqref{SaturationFaccio} suggests that the displacement $\Delta S(\mathbf{k}_{\perp})$  tends at small $k_{\perp}$ towards the intuitive geometric value given by the product of the sound velocity $c_s$ by the effective time $L$.

Surprisingly, this geometric approach differs drastically from the results of the full theory in~\cite{Larre2017,Ferreira2018} which predict instead a linear increase of $\Delta S(\mathbf{k}_{\perp})$ with $\Lambda =2\pi/k_{\perp}$ at small $k_{\perp}$, even in the limit of weak interactions ($\Delta n \to 0$). 
In the following, we explain the physical origin of this correction and show that interferences between the counter-propagating Bogoliubov collective excitations are responsible for the disagreement between \cite{Vocke2015} and \cite{Larre2017,Ferreira2018} in the sonic regime ($k_{\perp}\xi\ll1$).


\subsection{Theoretical model including the interferences between counter-propagating Bogoliubov excitations}
\label{SubSec:ActualTheory}

Let us first introduce qualitatively this effect before deriving the full analytical solution.
When the background and the probe enter the non-linear medium, both experience a sudden jump of the $\smash{\chi^{(3)}}$ susceptibility, analogous to a quantum quench of the interactions \cite{Larre2015, Larre2016, Larre2018}.
This generates a conjugate beam, due to the boundary condition on the electric field amplitude at the interface.
The conjugate field oscillates at the same frequency as the background and the probe, and propagates in the transverse direction with a wavevector $-\mathbf{k}_{\perp}$ opposite to the one of the incident probe.
In optics, this third-order non-linear wave-mixing process is known as degenerate four-wave mixing \cite{glorieux2010double,glorieux2012generation,agha2011time}. 

Interestingly, the interferences between the two counter-propagating Bogoliubov excitations (the probe and the conjugate within the medium), neglected in the geometrical model~\cite{Vocke2015}, are continuously taking place all along their propagation in the non-linear medium.
Since the pump, probe and conjugate have the same frequency, they do fulfill the phase-matching condition only when they are co-propagating, that is, when $k_{\perp}=0$.
This can be seen by evaluating the ratio of the conjugate Bogoliubov amplitude $v(\mathbf{k}_{\perp})$ to the probe one $u(\mathbf{k}_{\perp})$ using \eqref{BogAmplitudes}.
In the free-particle regime $k_{\perp}\xi\gg1$, this ratio is small as it scales as $1/(k_{\perp}^{2}\xi^{2})$. 
In this limit, the impact of the interferences between the conjugate and the probe can be safely neglected and \eqref{HPPhaseDiff} is valid, as shown in the next section. 
However, in the phonon regime $k_{\perp}\xi\ll1$, $|v(\mathbf{k}_{\perp})/u(\mathbf{k}_{\perp})|\simeq 1$ and a full model taking into account the interferences between the counter-propagating Bogoliubov excitations gives drastically different results from the geometric approach detailed above. 

In the case of a finite diameter probe mode (not a plane wave), the geometric model of \eqref{ShiftFaccio} is recovered when the length $L$ of the medium is long enough for the probe and conjugate wavepackets to get spatially separated during the propagation~\cite{carusotto2014superfluid}.
In this limit, the distance between of the wavepackets centers gives access to the group velocity~\cite{Fontaine2018}, while the position of the fringes within the wavepackets gives access to the phase velocity. For realistic parameters, this requires impractically long samples. 

In the following, we derive an exact expression for the relative phase $\Phi_{\mathrm{NL}}(\mathbf{k}_{\perp})=\Phi_{\mathrm{p},\mathrm{NL}}(\mathbf{k}_{\perp})-\Phi_{\mathrm{bg},\mathrm{NL}}$ accumulated by the probe with respect to the background after propagation through the medium.
We use an approach similar to the quantum optics input-ouput formalism \cite{reynaud1992quantum,courty1992generalized} with a description of the medium given by the Bogoliubov theory \cite{Larre2017}.

\noindent In air (i.e. $\smash{z<0}$ and $\smash{z>L}$) the envelope of the electric field including the background and its fluctuations may be expanded as
\begin{equation}
    \mathcal{E}_{\mathrm{air}}(\mathbf{r}_{\perp},z)=\sqrt{\rho_{\mathrm{air}}(z)}e^{i\Phi_{\mathrm{air}}(z)}+e^{i\Phi_{\mathrm{air}}(z)}\int\frac{d^{2}\mathbf{k}_{\perp}}{(2\pi)^{2}}a_{\mathbf{k}_{\perp}}(z)e^{i\mathbf{k}_{\perp}\cdot\mathbf{r}_{\perp}}.
    \label{eqAir}
\end{equation}
In this equation, $\rho_{\mathrm{air}}(z)$ and $\Phi_{\mathrm{air}}(z)$ are the density and the phase of the homogeneous background in air.
Due to the conservation of energy at $z=0$ and $z=L$, the densities are related by $\rho_{\mathrm{air}}=\rho_{\mathrm{air}}(z>L)=n\rho_{0}$, while the phases are $\Phi_{\mathrm{air}}(z<0)=0$, and $\Phi_{\mathrm{air}}(z>L)=-k_0\Delta nL$.
In \eqref{eqAir}, $a_{\mathbf{k}_{\perp}}(z)$ denotes the Fourier amplitude of the fluctuations superimposed on the background in air.
In our experiment, only one $\mathbf{k}_{\perp}$ component (corresponding to the probe mode for $z=0^-$) is injected into the medium and all the other modes are set to zero.
Using the sign convention adopted in \eqref{Fluctuation}, the phase difference $\Phi_{\mathrm{NL}}(\mathbf{k}_{\perp})$ can then be expressed as follows:
\begin{equation}
    \label{TruePhiNL}
    \Phi_{\mathrm{NL}}(\mathbf{k}_{\perp})=-\mathrm{arg}\bigg[\frac{a_{\mathbf{k}_{\perp}}(L^+)}{a_{\mathbf{k}_{\perp}}(0^-)}\bigg].
\end{equation}
To derive the input/output relation between $a_{\mathbf{k}_{\perp}}(L^+)$ and $a_{\mathbf{k}_{\perp}}(0^-)$ we need to use both energy conservation at the interfaces and the Bogoliubov formalism for the evolution within the medium.
In a first step, when entering the medium at $z=0$, the probe of amplitude $a_{\mathbf{k}_{\perp}}(0^-)$ transforms, by energy conservation, into $a_{\mathbf{k}_{\perp}}(0^+)=\sqrt{n}\,a_{\mathbf{k}_{\perp}}(0^-)$.
Then, in analogy to the quantum formalism of dilute Bose gases, we can consider the term $a_{\mathbf{k}_{\perp}}(0^+)$ to be equivalent to the annihilation operator for the weakly interacting particles.
Following the Bogoliubov approach, it can be connected to the non-interacting Bogoliubov operators $b_{\mathbf{k}_{\perp}}$ using the transformation:
\begin{align}
    \label{ContinuityAirMedium}
    \renewcommand\arraystretch{1}
    \begin{bmatrix}
        a_{\mathbf{k}_{\perp}}(0^+) \\
        a_{-\mathbf{k}_{\perp}}^{\ast}(0^+)
    \end{bmatrix}
    =
    \renewcommand\arraystretch{1}
    \begin{bmatrix}
        u(\mathbf{k}_{\perp}) & \!\!\! v(\mathbf{k}_{\perp}) \\
        v(\mathbf{k}_{\perp}) & \!\!\! u(\mathbf{k}_{\perp})
    \end{bmatrix}
    \renewcommand\arraystretch{1}
    \begin{bmatrix}
        b_{\mathbf{k}_{\perp}} \\
        b_{-\mathbf{k}_{\perp}}^{\ast}
    \end{bmatrix}.
\end{align}
Thereafter, the counter-propagating Bogoliubov excitations evolve along the optical axis, accumulating a propagation phase provided by the Bogoliubov dispersion relation $\Omega_{\mathrm{B}}(\mathbf{k}_{\perp})$.
This evolution is analogous to those observed after a quench in atomic BEC, and leads to synchronized phases between the counter-propagating phonon modes ($b_{\mathbf{k}_{\perp}}$ and $b_{-\mathbf{k}_{\perp}}^{\ast}$). Interestingly, this synchronized effect is at the origin of Sakharov oscillations \cite{hung2013cosmology}.
We obtain:
\begin{align}
 \label{ContinuityMediumAir}
    \renewcommand\arraystretch{1}
    \begin{bmatrix}
        a_{\mathbf{k}_{\perp}}(L^-) \\
        a_{-\mathbf{k}_{\perp}}^{\ast}(L^-)
    \end{bmatrix}
    =
    \renewcommand\arraystretch{1}
    \begin{bmatrix}
        u(\mathbf{k}_{\perp})e^{-i\Omega_{\mathrm{B}}(\mathbf{k}_{\perp})L} & \!\!\!
        v(\mathbf{k}_{\perp})e^{i\Omega_{\mathrm{B}}(\mathbf{k}_{\perp})L} \\
        v(\mathbf{k}_{\perp})e^{-i\Omega_{\mathrm{B}}(\mathbf{k}_{\perp})L} & \!\!\!
        u(\mathbf{k}_{\perp})e^{i\Omega_{\mathrm{B}}(\mathbf{k}_{\perp})L}
    \end{bmatrix}
    \renewcommand\arraystretch{1}
    \begin{bmatrix}
        b_{\mathbf{k}_{\perp}} \\
        b_{-\mathbf{k}_{\perp}}^{\ast}
    \end{bmatrix}.
\end{align}
We can then invert \eqref{ContinuityAirMedium} using the Bogoliubov normalization $u^{2}(\mathbf{k}_{\perp})-v^{2}(\mathbf{k}_{\perp})=1$ and inject the result into the right-hand side of \eqref{ContinuityMediumAir}.
Finally, taking into account energy conservation at the medium output $a_{\mathbf{k}_{\perp}}(L^-)=\sqrt{n}\,a_{\mathbf{k}_{\perp}}(L^+)$, we obtain the following input-output relations:
\begin{equation}
    \label{InputOutputRelation}
    \renewcommand\arraystretch{1}
    \begin{bmatrix}
        a_{\mathbf{k}_{\perp}}(L^+) \\
        a_{-\mathbf{k}_{\perp}}^{\ast}(L^+)
    \end{bmatrix}
    =
    \renewcommand\arraystretch{1}
    \begin{bmatrix}
        U(\mathbf{k}_{\perp}) & \!\!\! V^{\ast}(-\mathbf{k}_{\perp}) \\
        V(\mathbf{k}_{\perp}) & \!\!\! U^{\ast}(-\mathbf{k}_{\perp}) \\
    \end{bmatrix}
    \renewcommand\arraystretch{1}
    \begin{bmatrix}
        a_{\mathbf{k}_{\perp}}(0^-) \\
        a_{-\mathbf{k}_{\perp}}^{\ast}(0^-)
    \end{bmatrix}
    ,
\end{equation}
where $U(\mathbf{k}_{\perp})$ and $V(\mathbf{k}_{\perp})$ are defined by
\begin{align}
    \label{U}
    U(\mathbf{k}_{\perp})&=u^{2}(\mathbf{k}_{\perp})e^{-i\Omega_{\mathrm{B}}(\mathbf{k}_{\perp})L}-v^{2}(\mathbf{k}_{\perp})e^{i\Omega_{\mathrm{B}}(\mathbf{k}_{\perp})L}\quad\text{and} \\
    \label{V}
    V(\mathbf{k}_{\perp})&=-2iu(\mathbf{k}_{\perp})v(\mathbf{k}_{\perp})\sin[\Omega_{\mathrm{B}}(\mathbf{k}_{\perp})L].
\end{align}
In our configuration, right before the medium entrance ($z=0^-$), the mode with wavevector $-\mathbf{k}_{\perp}$ has a zero amplitude (i.e., the conjugate mode is seeded by vacuum) and therefore we set $a_{-\mathbf{k}_{\perp}}^{\ast}(0^-)=0$ in \eqref{InputOutputRelation}.
As a result, we eventually come to the simple relation
\begin{equation}
    \label{ProbeProbe}
    a_{\mathbf{k}_{\perp}}(L^+)=U(\mathbf{k}_{\perp})a_{\mathbf{k}_{\perp}}(0^-), 
\end{equation}
from which we can simplify \eqref{TruePhiNL} and obtain the $\Omega_{\rm B}(\mathbf{k}_{\perp})$ dependence of $\Phi_{\mathrm{NL}}(\mathbf{k}_{\perp})$ from Eqs.~(\ref{DispersionRelation}), (\ref{BogAmplitudes}), and (\ref{U}):
\begin{equation}
    \label{HPPhaseLossless}
    \begin{split}
    \Phi_{\mathrm{NL}}(\mathbf{k}_{\perp})&=-\mathrm{arg}[U(\mathbf{k}_{\perp})] \\
    &=\Omega_B(\mathbf{k}_{\perp}) L -\mathrm{arg}[u^2(\mathbf{k}_{\perp})-v^2(\mathbf{k}_{\perp}) e^{2i\Omega_B(\mathbf{k}_{\perp}) L}] \\
    &=\arctan\bigg\{\frac{[k_{\perp}^{2}/(2k_{0})]^{2}+\Omega_{\rm B}(\mathbf{k}_{\perp})^{2}}{k_{\perp}^{2}/k_{0}\times\Omega_{\rm B}(\mathbf{k}_{\perp})}\tan[\Omega_{\rm B}(\mathbf{k}_{\perp})L]\bigg\}.
\end{split}
\end{equation}



\noindent The second expression of \eqref{HPPhaseLossless} allows for a direct understanding of the role of the interferences between Bogoliubov phonon excitations in the correction to \eqref{ShiftFaccio}.

In the free-particle regime ($k_{\perp}\xi\gg1$),
the $v^2$ term in the second expression of \eqref{HPPhaseLossless} is negligible. This is equivalent to say that the phase-matching condition is not fulfilled and the four-wave-mixing process is inefficient to create the conjugate mode.
Since $u(\mathbf{k}_{\perp})$ is real, we get
\begin{equation}
    \Phi_{\mathrm{NL}}(\mathbf{k}_{\perp})\underset{k_{\perp}\xi\gg1}{\simeq} \Omega_{\rm B}(\mathbf{k}_{\perp}) L.
\end{equation}


\begin{figure}[t!]
\centering
\includegraphics[width=\columnwidth]{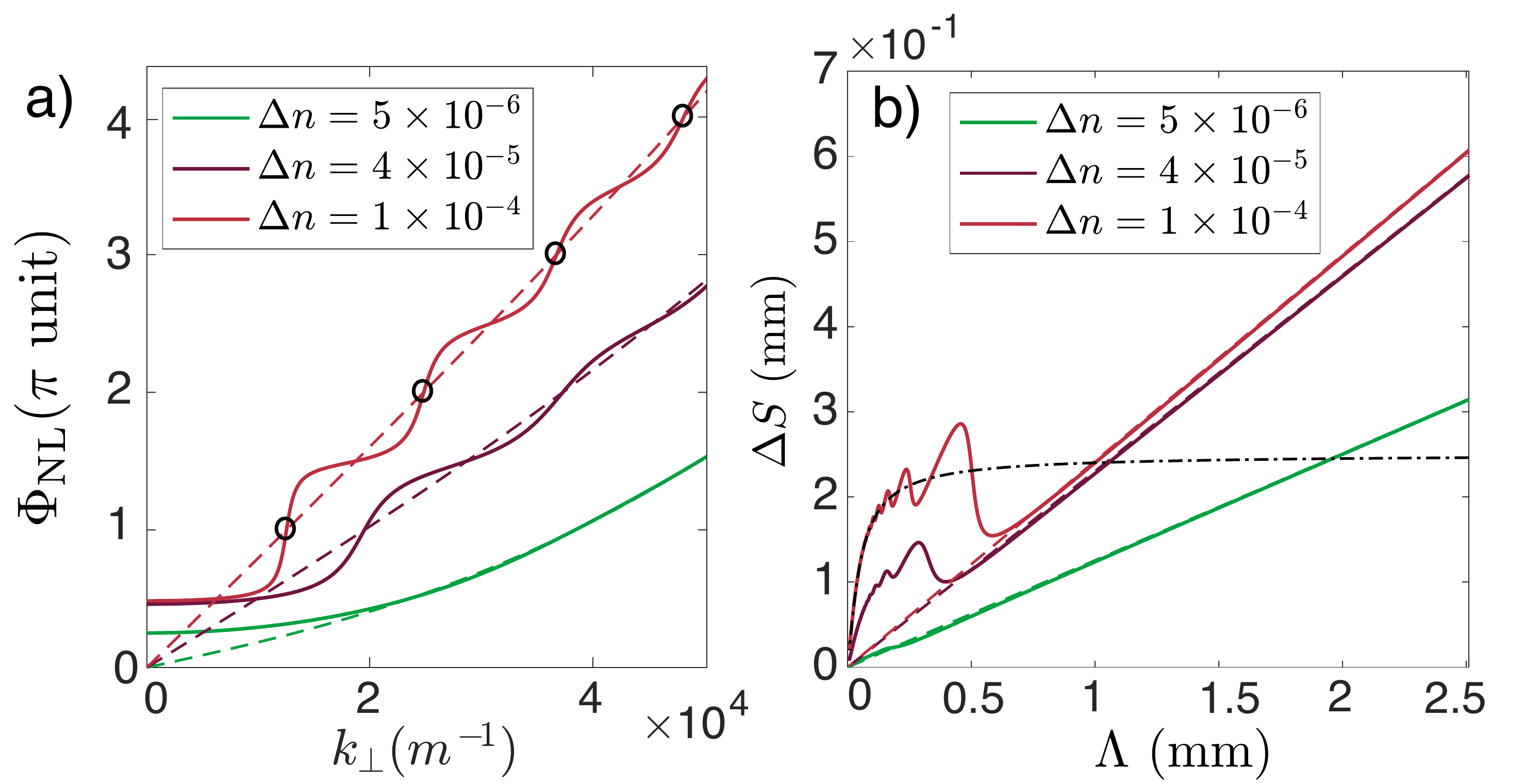}
\caption{(a) Non-linear phase shift $\Phi_{\mathrm{NL}}(\mathbf{k}_{\perp})$ and (b) relative fringes displacement $\Delta S(\Lambda)$ as functions of, respectively, the Bogoliubov wavenumber $k_{\perp}=|\mathbf{k}_{\perp}|$ and the Bogoliubov wavelength $\Lambda = 2 \pi / k_{\perp}$ for different values of the optical non-linearity $\Delta n$. 
(a) The phase shift $\Phi_{\mathrm{NL}}(\mathbf{k}_{\perp})$ generally follows $\Omega_{\rm B}(\mathbf{k_{\perp}}) L$ in average (dashed lines), except for small $k_{\perp}$,  where it saturates at a non-zero value for all values of $\Delta n \neq 0$.
According to \eqref{LinearTrend}, the ${k}_{\perp}=0$ limit  $\Phi_{\mathrm{NL}}(\mathbf{k}_{\perp}=0)$ is a growing function of $\Delta n$ and tends towards $\pi/2$ for large $\Delta n$.
(b) The staircase structure of $\Phi_{\mathrm{NL}}(\mathbf{k}_{\perp})$ translates into oscillations in $\Delta S(\Lambda)$, most visible for large interactions.
At long $\Lambda$, $\Delta S(\Lambda)$ increases linearly according to \eqref{ShiftLinearTrend}. This trend is present even for weak interactions  ($\Delta n \to 0$).
Solid lines corresponds to the full model and the black dashed-dotted line shows for comparison the displacement obtained from the geometric approach of \eqref{ShiftFaccio} with $\Delta n = 10^{-4}$.}
\label{fig:ThPhaseShift}
\end{figure}

\noindent This limit exactly corresponds to the geometric model of \eqref{HPPhaseDiff} and correctly describes the transverse fringes displacement $\Delta S(\mathbf{k}_{\perp})$. 
However, this approximation is only valid in the parabolic dispersion limit at large momenta $k_{\perp}\xi\gg1$.

In the phonon regime ($k_{\perp} \xi \ll 1$) where superfluidity is manifest, there are fundamental differences between the predictions of the geometrical approach and the full model, because the $v^2$ term cannot be neglected anymore in the second expression of \eqref{HPPhaseLossless}.
This interference term leads to a correction to \eqref{HPPhaseDiff}, which we can expand analytically in the limit $k_{\perp}\xi\ll1$ to get at the leading order
\begin{equation}
     \Phi_{\mathrm{NL}}(\mathbf{k}_{\perp})\underset{k_{\perp}\xi\ll1}{=}\arctan(2k_{0}\Delta nL)+O(k_{\perp}^{2}\xi^{2}).
     \label{LinearTrend}
\end{equation}
An essential feature of \eqref{LinearTrend} is that the non-linear phase difference $\Phi_{\mathrm{NL}}(\mathbf{k}_{\perp})$ at the medium output converges towards a constant non-zero value for small $\mathbf{k}_{\perp}$.
This can be clearly seen in Fig.~\ref{fig:ThPhaseShift}(a) for realistic experimental parameters.
This non-zero value holds independently of the strength of the interactions $\Delta n$, and therefore is a general feature of paraxial fluids of light and a direct consequence of the interferences between Bogoliubov phonons.
At large $\Delta n$, this offset saturates towards $\pi/2$.

In between these two asymptotic limits, we also observe numerically a smooth staircase structure, which follows on average the trend of the geometric prediction \eqref{ShiftFaccio} in the large-$k_{\perp}$ limit (dashed lines in Fig.~\ref{fig:ThPhaseShift}(a)).
This staircase structure becomes more and more visible as the optical non-linearity $\Delta n$ increases.
This effect is less robust than the non-zero value of $\Phi_{\mathrm{NL}}(\mathbf{k}_{\perp})$ in the small $k_{\perp}$ limit previously described and does not hold for weak interactions $\Delta n$ (see the green curve of Fig.~\ref{fig:ThPhaseShift} (a)).
Therefore, in order to evidence the presence of interferences between the Bogoliubov phonons, we will focus our attention on the phase difference at small $k_{\perp}$ by looking at the displacement $\Delta S$ as function of $\Lambda = 2\pi/k_{\perp}$.



In Fig.~\ref{fig:ThPhaseShift}(b), we present this displacement $\Delta S$ (accessible experimentally) as function of the density modulation wavelength $\Lambda$.
Because of the staircase structure of $\Phi_{\mathrm{NL}}(\mathbf{k}_{\perp})$, the displacement $\Delta S(\Lambda)$ oscillates at short $\Lambda$.
Once again this effect disappear for weak interactions $\Delta n$ (green curve of Fig.~\ref{fig:ThPhaseShift} (b)).
In the contrary, the linear increase of $\Delta S(\Lambda)$ when $\Lambda\gg \xi $ is always present for all $\Delta n$ and can  be computed from \eqref{LinearTrend} as:
\begin{equation}
    \Delta S(\Lambda)\underset{\Lambda/\xi\gg1}{\simeq}\frac{\arctan(2k_{0}\Delta nL)}{2\pi}\Lambda,
    \label{ShiftLinearTrend}
\end{equation}
This expression significantly contrasts with \eqref{SaturationFaccio}, obtained within the geometrical approach.
For comparison, the total displacement (\eqref{ShiftFaccio}) predicted by the geometric method is plotted for $\Delta n=10^{-4}$ in black dashed-dotted line in Fig.~\ref{fig:ThPhaseShift}(b).
As expected, the two descriptions match in the free-particle regime, but  the  linear increase (\eqref{ShiftLinearTrend}) at long $\Lambda$ is only present in the full model and not predicted by \eqref{ShiftFaccio}.

In the next section, we explore experimentally this configuration in a hot atomic vapor to compare and verify the predictions of the two models.
We will show evidences of the interference between the counter-propagating Bogoliubov collective excitations at small ${k}_{\perp}$ and of their role on the measurement of the dispersion relation following \eqref{ShiftLinearTrend}.

\section{Experimental evidences of interferences between Bogoliubov excitations }
\begin{figure}[ht]
\center
\includegraphics[width=0.96\columnwidth]{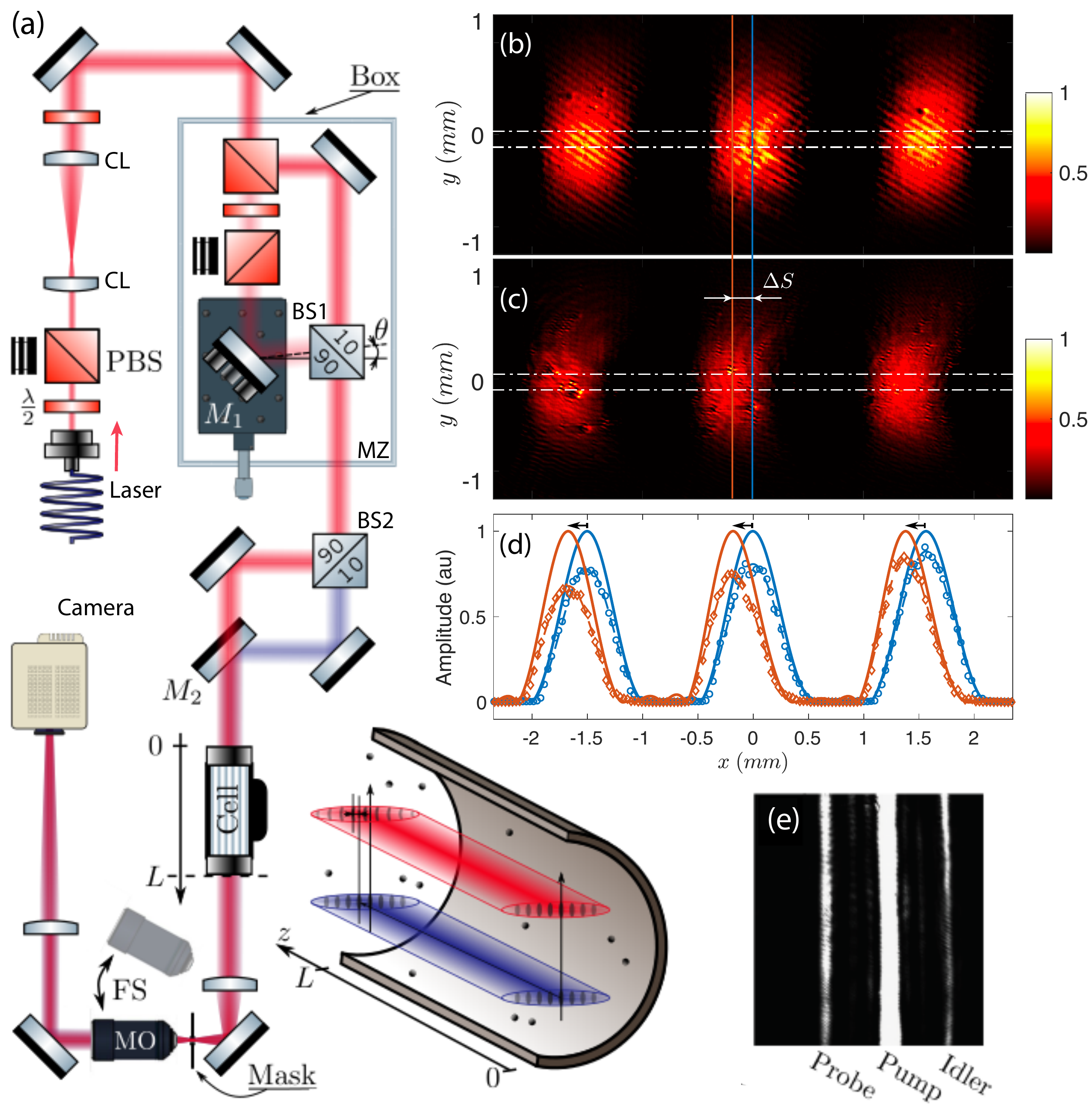} 
\caption{(a) Experimental setup. A laser is shaped with two cylindrical lenses (CL). It is split and recombined (BS1) within an unbalanced Mach-Zehnder (MZ) interferometer to create a low contrast fringes pattern. Two sets of fringes (low and high intensity) are vertically shifted using an 90:10 beam-splitter (BS2)   before going into an atomic vapor cell. The cell output is imaged on a camera after filtering in the Fourier space (FS).
(b-c) Background-subtracted images reveal the small amplitude density modulation which propagates on a low (b) and a high intensity background fluid (c).
The blue and red points in (d) are obtained by integrating the intensity in between the white dashed lines in (b) and (c) respectively. 
We first filter out the high frequency noise (dashed lines) and then normalize the envelopes (solid lines).
The shift is computed by measuring the nearest peak-to-peak distance between the solid lines.
(e) Fourier space image obtained by inserting a microscope objective (MO).} 
\label{fig:ShiftExp}
\end{figure} 
\subsection{Experimental setup}
Our experimental setup is sketched in figure~\ref{fig:ShiftExp} (a).
A continuous-wave laser field at $780$~nm is  elongated in the $x$ direction using a set of two cylindrical lenses.
This cylindrical telescope is slightly defocused in order to loosely focus the beam onto the medium input facet.
In this plane, the minor axis width$\;\omega_{0,y}$ (radius at $1/e^{2}$) is  $500$ $\mu$m while the major axis one, $\omega_{0,x}$ is  $1$ cm.
 The Rayleigh length associated to $\omega_{0,y}$ is much longer than the cell length ($L=7.5$~cm).
The cell is filled with an isotopically pure $^{85}$Rb vapor heated up to $400$ K. 
 The laser frequency is 2.6 GHz red-detuned with respect to the $F=3 \rightarrow F'$ transition of the $^{85}$Rb $D_{2}$ line, which ensure an linear index of refraction close to 1 and a transmission larger than 60~$\%$. 
 
The weak intensity modulation pattern is created using an unbalanced Mach-Zehnder interferometer.
The beam is then split in two with a $90\!:\!10$ ($R\!:\!T$) beam splitter and recombined with a vertical shift to have simultaneously a weak intensity modulation evolving on-top of a high intensity beam forming the photon fluid.
The medium exit plane is imaged on a CMOS camera with a $4f$ telescope.
By inserting a microscope objective on the beam path, we can image the momentum distribution (inset (e) of figure~\ref{fig:ShiftExp}).
Spatial Fourier filtering using a razor blade is conducted in this plane to filter out the conjugate beam that blurs the fringes pattern.
As sketched in figure~\ref{fig:ShiftExp}, we perform simultaneously the experiment in two regimes: (i) low fluid density and (ii) high fluid density. The low density fluid corresponds to the case of a negligible  non-linearity  and provides a reference ($\Phi_{\mathrm{L}}$) for the fringe displacement.
Comparing both patterns we observe the fringe displacement and measure $\Delta S$.

\subsection{Data analysis and results}

\begin{figure}[]
\center
\includegraphics[width=0.8\columnwidth]{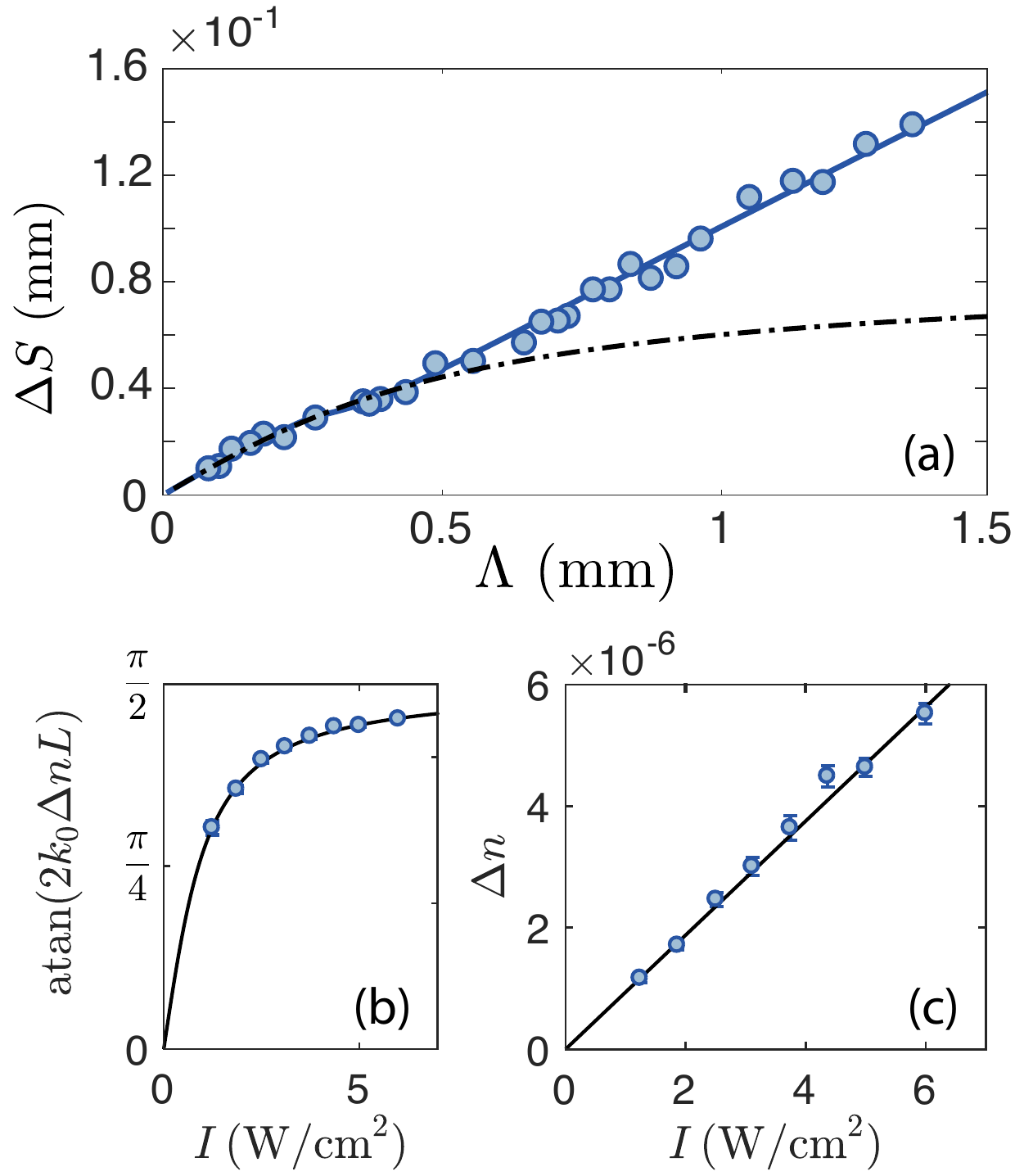} 
\caption{(a) Displacement $\Delta S$ as function of the modulation wavelength $\Lambda$ for a fluid intensity of $1.3$~W.cm$^{-2}$. 
The laser is 2.6 GHz red-detuned with respect to the $F=3 \rightarrow F'$ transition of the $^{85}$Rb $D_{2}$ line and the cell length is 7.5~cm.
The experimental data (blue circle) are fitted with the full theory (blue line) for $\Delta n = 1.3 \, 10^{-6}$. 
For comparison, the displacement obtained using~\eqref{ShiftFaccio} has been plotted (black dashed dotted line). 
(b) Slope of the asymptotic linear increase of $\Delta S$ at large $\Lambda$ as function of the fluid intensity. (c) $\Delta n$ extracted with eq.~\eqref{LinearTrend} from the slope of $\Delta S$ at large $\Lambda$ as function of the fluid intensity. The linear scaling of $\Delta n$ with $I$ confirms that we are not saturating the non-linearity, i.e. $\Delta n =n_2 I$ with $n_2=1\times10^{-10}$m$^2$/W.
}
\label{fig:MainShiftRes}
\end{figure}

After removing the background intensity distribution to keep only the small density modulation on top of it, typical interference patterns obtained at the medium output plane are shown in figures~\ref{fig:ShiftExp} (b-c).
 The displacement between the fringes of the low intensity reference (b) and high intensity fluid (c) $\Delta S$ is clearly visible.
We can note that the fringes are slightly bent in (c) because the intensity profile along the vertical axis is Gaussian and therefore the nonlinear phase shift accumulated during the propagation depends on $y$.  
In order to avoid errors during the data analysis, we average the intensity profile over the central region in between the white dotted line in figure~\ref{fig:ShiftExp} (b-c). 

After averaging, the resulting profiles are plotted in the (d) panel of Fig.~\ref{fig:ShiftExp}: the blue points are for the low intensity reference (a) while the red ones are for the high intensity non-linear case (b).
The high frequency noise is filtered out and we remove the envelopes using a cubic spline interpolation method to normalize it and obtain the blue and red solid curves. 
The relative displacement is computed by averaging on several fringes the distance  (black arrows) between the nearest maxima in the low intensity reference and in the high intensity case.

In figure~\ref{fig:MainShiftRes} (a), we present the experimentally measured $\Delta S$ as function of the modulation wavelength $\Lambda$. 
 The probe power is taken to ensure a modulation depth of less than $5\%$. 
 The full model is shown in blue solid line.
 For comparison, the geometrical model computed with~\eqref{ShiftFaccio} is plotted in black dashed dotted line.
These experimental results are a clear evidence that the geometrical model fails to describe the displacement $\Delta S$ at large $\Lambda$. 
Indeed, at large $\Lambda$, we observe a clear signature of the linear increase of $\Delta S$, as predicted by the full model.
By including the interferences between elementary Bogoliubov excitations, the full model also allows to predict the value of the slope as function of the nonlinear refractive index change $\Delta n$.  
To verify the consistency of our model, we repeated the measurement of $\Delta S$ for various field intensities $I$ and estimated $\Delta n$ from the theoretical predictions, using \eqref{ShiftLinearTrend}.
An intriguing feature of this equation is the  non-linear behavior of the phase shift and the saturation at large interaction $\Delta n$ (figure~\ref{fig:MainShiftRes} (b)).
However, as visible in figure~\ref{fig:MainShiftRes} (c),  the value of $\Delta n$ extracted from \eqref{ShiftLinearTrend}, depends linearly with the background intensity $I$ as expected for a Kerr medium and it  validates our experimental approach. 


\section{Interferences between Bogoliubov waves}
\subsection{Is this interference effect robust with respect to corrections to the lossless local Kerr model ?}

Several non-linear media have been proposed and implemented for fluid of light experiments, including atomic vapor \cite{vsantic2018nonequilibrium, fontaine2019attenuation}, methanol \cite{vocke2016role}, photo-refractive crystal \cite{wan2007dispersive, michel2018superfluid, boughdad2019anisotropic} and nematic liquid crystals \cite{Ferreira2018}. 
In these systems the microscopic origin of light-matter interaction strongly differs and can impact the properties of these fluids of light.
To verify that these variations do not change significantly the long wavelength behavior of $\Delta S$, we numerically studied the dependence on four key parameters: (i) the losses $\alpha$, (ii) the width of the pump beam  $w_{0,y}$, (iii) the non-locality and the (iv) saturation of the non-linear response.

\begin{figure}[ht!]
\center
\includegraphics[width=0.88\columnwidth]{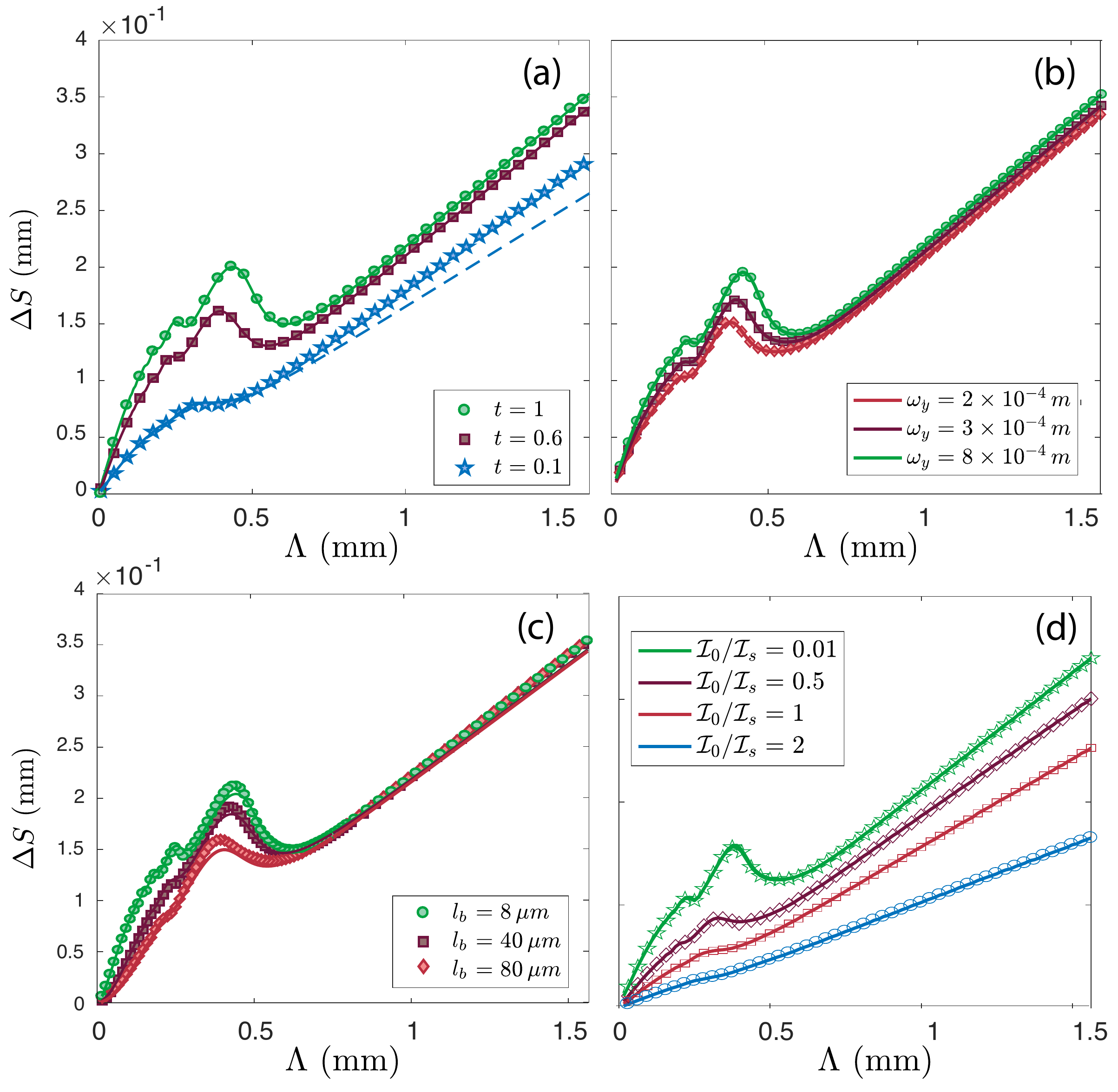} 
\caption{Numerical simulations (symbols) and analytical solutions (solid and dashed line) of $\Delta S$ as function of $\Lambda$.
(a) Different cell transmissions $t$. The simulations and the theory model are similar as long as the transmission remains large ($t > 0.5$). 
(b) Different background widths $w_{0,y}$. 
(c) Different non-local transport length scales $l_{b}$.
The oscillations are smoothed by non-locality.
In our system $l_b<10\mu$m.
(d) Different saturation intensity  $\mathcal{I}_s$ of the non-linear Kerr interaction. Once again oscillations are smoothed by a saturation of the medium.
For all the simulations $\Delta n=1.0 \, 10^{-5}$.}  
\label{fig:ShiftVersus}
\end{figure}

All the simulations have been performed using using a second order split-step method on the 2D nonlinear Schr\"{o}dinger equation and  a common set of parameters.
The background intensity is set to $\rho_0=2.5 \, 10^{5}$ W/m$^{2}$, the linear index to $n=1$, and the nonlinear index to $n_2=4 \, 10^{-11}$ m$^{2}$/W.
In the lossless situation, the nonlinear change of refractive index is  equal to $\Delta n=1.0 \, 10^{-5}$. 
The simulation results are presented in figure~\ref{fig:ShiftVersus}.

In figure~\ref{fig:ShiftVersus} (a), the displacement $\Delta S$ is plotted, for different cell transmissions $t = \exp(-\alpha L)$.
The colored points stem from numerical simulations whereas the theoretical curves are plotted in black solid.
A full derivation of the analytical model is given in the Supplementary Materials.
Absorption smooths out the oscillations at small $\Lambda$, similarly to a reduction of the non-linear interactions $\Delta n$ as seen in Fig.\ref{fig:ThPhaseShift}.
However the long-$\Lambda$ limit is qualitatively unchanged from the lossless case.
The analytical predictions (dashed lines) gives a accurate estimation of the long-$\Lambda$ slope for transmission larger than 0.5.

In figure~\ref{fig:ShiftVersus} (b), the effect of the finite beam width $w_{0,y}$ on the displacement $\Delta S$ is studied. 
We notice a reduction in the displacement oscillations amplitude when $w_{0,y}$ decreases.
But as for the absorption, this effect does not affect the general shape of the displacement curve and its large $\Lambda$ linear trend.
It can be understood intuitively, because for smaller beam width $w_{0,y}$, the Kerr self defocusing effect increases and therefore the background density spreads faster in the transverse plane along the propagation. 
This results in a decrease of the beam intensity on the major axis during the propagation and a consequent reduction of the effective interaction $\Delta n$. 

In figure~\ref{fig:ShiftVersus} (c), the impact of non-locality is reported.
The nonlinear phase shift formula~\eqref{HPPhaseDiff} has been generalized using the non-local dispersion relation to take ballistic transport of excited atoms into account in the theory (see supplementary materials for details).
 The theoretical predictions are plotted in black solid and match perfectly with simulations.
The main effect here is more subtle than the ones of the losses or the finite width of the beam.
The slope of the linear trend at high $\Lambda$ remains unchanged but a significant modification of the displacement in the oscillating part is observed.
This effect becomes significant for non-local ballistic length scales $l_{b}$ much longer than the typical ones of atomic vapors (typically, $l_{d} \approx 8$ $\mu$m at 400 K). 
The situation is very different in the thermo-optic media considered in~\cite{Vocke2015}, where the non-local length is on the order of 100~$\mu$m \cite{vocke2016role} and thus is able to significantly modify the behavior of the displacement for small $\Lambda$. 

Finally, in figure~\ref{fig:ShiftVersus} (d), we have studied the impact of a saturation of the non-linearity. 
The interaction strength $\Delta_n$ is replaced by $\Delta n\times \frac{1}{1+I/\mathcal{I}_s}$, where $\mathcal{I}_s$ is the saturation intensity.
This model reproduces saturation observed in atomic media and photorefractive crystals.
Compared to losses, the finite beam width, and non-locality, the effect of saturation on the displacement is the most important, as it not only attenuates the oscillations at small $\Lambda$ but also modifies the slope at large $\Lambda$.
This correction is a consequence of the reduction of the sound velocity by a larger factor to $c_s\times\frac{1}{(1+I/\mathcal{I}_s)^2}$. 
Nevertheless, saturation does not affect the large-$\Lambda$ behavior  of the displacement and, in particular, does not lead to the constant value for the large-$\Lambda$ limit predicted by the geometrical approach.

All these simulations confirm that the corrections to the ideal lossless model are able to modify the behavior of $\Delta S$ at small $\Lambda$, but do not affect the linear trend at large $\Lambda$.
The impact of the interferences between Bogoliubov modes is therefore robust and can thus be envisioned as a novel tool to probe the dispersion and the static structure factor of the photon fluid, in a similar way to what was done with atomic BEC \cite{shammass2012phonon}.
In the last part of this work, we propose an explanation for the robustness of these interferences and for their importance to understand the superfluid behavior based on a universal mechanism known as the Sakharov oscillations \cite{sakharov1966initial,hung2013cosmology}.


\vspace{-0.5cm}

\subsection{Stimulated Sakharov-like oscillations}



The Bogoliubov excitation (\eqref{Fluctuation}) generated at the entrance of the non-linear medium consists in a superposition of counter-propagating plane-waves in the $\mathbf{r}_{\perp}$ plane with opposite wavevectors $\mathbf{k}_{\perp}$ and $-\mathbf{k}_{\perp}$.
These Bogoliubov components are simultaneously generated at the medium entrance and oscillate at the respective angular frequencies $\Omega_{\rm B}(\mathbf{k}_{\perp})$ and $-\Omega_{\rm B}(\mathbf{k}_{\perp})$ along the propagation axis, which is analogous to time.
As a consequence, at a given effective time $z$, these components will have acquired a relative phase difference of $2\Omega_{\rm B}(\mathbf{k}_{\perp})z$.
Interestingly, this behavior is very similar to the one predicted for the Sakharov oscillations in cosmology \cite{sakharov1966initial,cosmobook} and can be understood in terms of the interference between the counter-propagating phonons that are spontaneously generated after a quantum quench \cite{hung2013cosmology, Martone2018, Robertson2017}.
Here, we draw the analogy and we  consider our experimental observations as a stimulated analogue of the Sakharov-like oscillations by seeding phonons on the $+\mathbf{k}_{\perp}$ mode.

\begin{figure}[t!]
\centering
\includegraphics[width=0.95\columnwidth]{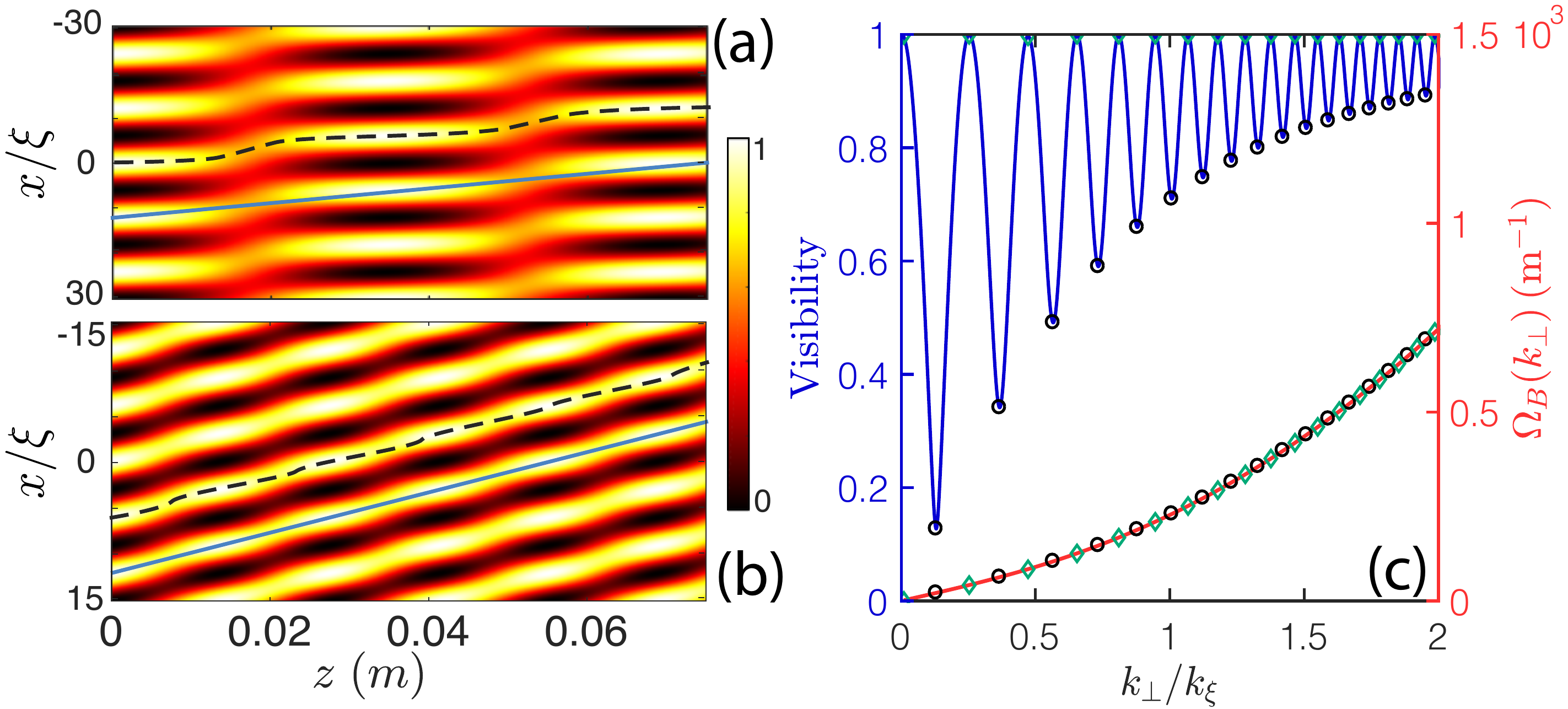}
\caption{Evolution along the $z$ axis of the transverse field intensity in a given $y$ plane for (a) $k_{\perp}\xi=0.5$ and (b) $k_{\perp}\xi=1$. 
The background intensity is subtracted on both images.
The black dashed curves follow the center of a bright fringe. 
The blue solid line is a trajectory of a Bogoliubov mode at the speed of sound.
(c) Visibility of the interference fringes at the output plane $z=L$ as function of $k_{\perp}$ (solid blue line). Visibility maxima are shown by green diamonds and minima are shown by black circle.
The dispersion relation (solid red line - right axis) is reconstructed using a sampling based on the position of the maxima ($\Omega_{\rm B}(\mathbf{k}_{\perp})  = p \pi/L$) and minima ($\Omega_{\rm B}(\mathbf{k}_{\perp})  = (p+1/2) \pi/L$).
Here $\Delta n=1.0 \, 10^{-5}$.}  
\label{fig:Contrast}
\end{figure}

For paraxial fluids of light experiments, we only have access  to the intensity at $z=L$ and not inside the medium.
Therefore, we have solved numerically the nonlinear Schr\"{o}dinger equation (\eqref{NLSE}) and computed the intensity of the total electric field inside the non-linear medium at every transverse planes along the $z$ axis to evidence the stimulated Sakharov-like oscillations in this optical system.
In figure~\ref{fig:Contrast}, we present the intensity profiles along $x$ in both the phonon regime ($k_{\perp}\xi<1$) in panel (a) and in the free-particle regime ($k_{\perp}\xi\leq1$) in panel (b).
On this figure, the background fluid density has been subtracted.
Two remarkable observations can be highlighted in figure~\ref{fig:Contrast} . 

First, constructive (maximum contrast) or destructive (minimum contrast) interferences between the counter-propagating Bogoliubov waves are clearly visible in the transverse direction. 
Along $z$, constructive interferences are located at $\Omega_{\rm B}(\mathbf{k}_{\perp}) z = p \pi$, with $p\geqslant1$ integer valued.
Reversely, when $\Omega_{\rm B}(\mathbf{k}_{\perp}) z = (p+1/2) \pi$, Bogoliubov modes destructively interfere and the contrast is minimum. 
Interestingly, these interference patterns can also be observed at a fixed effective time (e.g. $z=L$) by changing the value of $\mathbf{k}_{\perp}$.
In figure~\ref{fig:Contrast} (c), we have extracted the dispersion relation using this approach.
At fixed $z=L$, we reported each value of $\mathbf{k}_{\perp}$ leading to a visibility maximum, as we know that  $\Omega_{\rm B}(\mathbf{k}_{\perp})  = p \pi/L$ (green diamonds on figure~\ref{fig:Contrast} (c)).
To increase the resolution of the reconstruction we can apply the same procedure with the visibility minima (black circles on figure~\ref{fig:Contrast}(c)) and obtain a sampling of the dispersion for $\Omega_{\rm B}(\mathbf{k}_{\perp})  = (p+1/2) \pi/L$.


Secondly, we can notice that the reduction in the contrast of the interference fringes that is observed when the two Bogoliubov components destructively interfere is more pronounced at low (Fig. \ref{fig:Contrast} (a)) than at large (Fig. \ref{fig:Contrast} (b)) wavevectors. This effect is not present in the spontaneous Sakharov oscillations triggered by zero-point fluctuations \cite{hung2013cosmology} and is a direct consequence of the stimulation of the process by the classical incident field in the $+\mathbf{k}_{\perp}$ mode. 

Indeed by seeding the process we break the symmetry between $+\mathbf{k}_{\perp}$ and $-\mathbf{k}_{\perp}$ modes, so the visibility reduction can be understood  by comparing $|u^2(\mathbf{k}_{\perp})|$ and $|v^2(\mathbf{k}_{\perp})|$ using \eqref{DispersionRelation}.
When $k_{\perp}\xi\gg 1$, then $|v^2(\mathbf{k}_{\perp})|$ becomes small comparing to $|u^2(\mathbf{k}_{\perp})|$ and therefore the interference contrast is reduced.
As a consequence, we can see in figure~\ref{fig:Contrast} that the trajectories of a bright fringe (black dashed line) are much less deformed with respect to the speed of sound propagation (blue solid line) for $k_{\perp}\xi=1$ (panel (b)) than for $k_{\perp}\xi= 0.5$ (panel (a)), where a staircase-like structure is apparent.
This exemplifies once again why the geometrical approach is a good approximation only in the free particle regime ($k_{\perp}\xi> 1$).

\section*{Conclusion}
In this work, we have studied the Bogoliubov excitations of a photon superfluid.
We have experimentally demonstrated a previously undetected phenomenon whereby the propagation of plane wave excitations in the fluid does not tend to the geometric prediction for the displacement, namely the product of the sound velocity $c_s$ by the effective time $L$, but keeps growing linearly with the excitation wavelength.
This is shown to be a direct consequence of the interference between counter-propagating Bogoliubov modes that are generated at an interaction quench and have only been observed in atomic superfluids \cite{cheneau2012light}.
These interferences can also be interpreted as stimulated Sakharov oscillations \cite{hung2013cosmology}, i.e. an analogue of fluctuations imprinted in the primordial Universe and visible as oscillations in the cosmic microwave background power spectrum \cite{sakharov1966initial,cosmobook}.
These results shows that these interferences are an essential element to describe accurately the dynamics of excitations in superfluids of light.
It brings a novel understanding of superfluidity for paraxial fluids of light and opens exciting perspectives for studying quantum effects in these systems, including quantum depletion and entanglement of phonons in Sakharov oscillations.

\vspace{-0.5cm}

\section*{Funding Information}
This work has received funding from the French ANR grant (C-FLigHT 138678, QFL) and from the European Union’s Horizon 2020 Research and Innovation Program under grant agreement No 820392 (PhoQuS). QG and AB thank the Institut Universitaire de France (IUF) for support. IC acknowledges support from the Provincia Autonoma di Trento.


%
\section*{Supplementary Materials}

\subsection*{Experimental alignment procedure}
In order to accurately measure the displacement $\Delta S$, one needs to precisely align the reference beam with respect to the high power one. The alignment procedure is as follow:
\begin{itemize}
    \item[(1)] We first make sure that both background beams (probe off) roughly propagate with the same transverse wave-vector and are correctly positioned one above the other (their respective center should lie on the same vertical axis).
    \item[(2)] We then switch the probe beam on. The next step is to align the interference fringes of the lower and upper interference patterns. We start by removing the cell and make sure that bright fringes on the bottom face bright fringes on the top. Of course, by doing so, the optical axis of the lower and upper beams are not parallel anymore. We should then switch to k-space, bring back the backgrounds to the initial position ($k_{\perp} = 0$) and repeat this procedure iteratively (beam walking). We finally check that for every transverse wave-vector $k_{\perp}$ the interference fringes remain aligned before putting the cell back on the beams path.
\end{itemize}

\subsection*{Photon absorption}

Photon absorption is described in \eqref{NLSE} by the term proportional to $\alpha\geqslant0$. When $\alpha\neq0$, the $\mathbf{r}_{\perp}$-independent electric-field envelope $\mathcal{E}_{0}$ and its linearized fluctuations $\delta\mathcal{E}$ acquire the following $z$-dependences:
\begin{align}
    \label{BackgroundAbsorption}
    \mathcal{E}_{0}(z)&\left.=\sqrt{\rho_{0}}e^{-\alpha z/2-ik_{0}\Delta n(1-e^{-\alpha z})/\alpha},\right. \\
    \notag
    \delta\mathcal{E}(\mathbf{r}_{\perp},z)&\left.=e^{-ik_{0}\Delta n(1-e^{-\alpha z})/\alpha}\right. \\
    \notag
    &\left.\hphantom{=}\times\int\frac{d^{2}\mathbf{k}_{\perp}}{(2\pi)^{2}}\Big\{u(\mathbf{k}_{\perp},z)b_{\mathbf{k}_{\perp}}e^{i[\mathbf{k}_{\perp}\cdot\mathbf{r}_{\perp}-\int_{0}^{z}dz'\Omega_{\mathrm{B}}(\mathbf{k}_{\perp},z')]}\right. \\
    \label{FluctuationsAbsorption}
    &\left.\hphantom{=}+v^{\ast}(\mathbf{k}_{\perp},z)b_{\mathbf{k}_{\perp}}^{\ast}e^{-i[\mathbf{k}_{\perp}\cdot\mathbf{r}_{\perp}-\int_{0}^{z}dz'\Omega_{\mathrm{B}}^{\ast}(\mathbf{k}_{\perp},z')]}\Big\}.\right.
\end{align}
In Eqs.~(\ref{BackgroundAbsorption}) and (\ref{FluctuationsAbsorption}), $\rho_{0}$ is the density of the paraxial fluid of light at $z=0$ and $\Delta n=g\rho_{0}/k_{0}$ is the corresponding non-linearity. We treat the $z$-dependence of the Bogoliubov spectrum $\Omega_{\mathrm{B}}$ and of the Bogoliubov amplitudes $u$ and $v$ in the adiabatic-evolution approximation \cite{born1928beweis, Larre2017}. Searching for real-valued $u$ and $v$ such that $u^{2}-v^{2}=1$ for all $z$, this gives
\begin{align}
    \label{BogoliubovSpectrumAbsorption}
    \Omega_{\mathrm{B}}(\mathbf{k}_{\perp},z)&=\sqrt{\frac{k_{\perp}^{2}}{2k_{0}}\bigg(\frac{k_{\perp}^{2}}{2k_{0}}+2k_{0}\Delta ne^{-\alpha z}\bigg)}-\frac{i\alpha}{2}, \\
    \label{BogoliubovAmplitudesAbsorption}
    u(\mathbf{k}_{\perp},z)\pm v(\mathbf{k}_{\perp},z)&=\bigg\{\frac{k_{\perp}^{2}}{2k_{0}}\bigg/\mathrm{Re}[\Omega_{\mathrm{B}}(\mathbf{k}_{\perp},z)]\bigg\}^{\pm\frac{1}{2}}.
\end{align}

All the observables computed in this paper rely on the input-output relation (\ref{InputOutputRelation}), which also holds when $\alpha\neq0$ provided (\ref{U}) and (\ref{V}) are respectively replaced with
\begin{align}
    \notag
    U(\mathbf{k}_{\perp})&\left.=u(\mathbf{k}_{\perp},0)u(\mathbf{k}_{\perp},L)e^{-i\int_{0}^{L}dz\Omega_{\mathrm{B}}(\mathbf{k}_{\perp},z)}\right. \\
    \label{UAbsorption}
    &\left.\hphantom{=}-v(\mathbf{k}_{\perp},0)v(\mathbf{k}_{\perp},L)e^{i\int_{0}^{L}dz\Omega_{\mathrm{B}}^{\ast}(\mathbf{k}_{\perp},z)},\right. \\
    \notag
    V(\mathbf{k}_{\perp})&\left.=u(\mathbf{k}_{\perp},0)v(\mathbf{k}_{\perp},L)e^{-i\int_{0}^{L}dz\Omega_{\mathrm{B}}(\mathbf{k}_{\perp},z)}\right. \\
    \label{VAbsorption}
    &\left.\hphantom{=}-v(\mathbf{k}_{\perp},0)u(\mathbf{k}_{\perp},L)e^{i\int_{0}^{L}dz\Omega_{\mathrm{B}}^{\ast}(\mathbf{k}_{\perp},z)}.\right.
\end{align}
For example, the non-linear phase $\Phi_{\mathrm{NL}}(\mathbf{k}_{\perp})$ expected for $\alpha\neq0$ reads
\begin{align}
    \notag
    &\left.\Phi_{\mathrm{NL}}(\mathbf{k}_{\perp})\right. \\
    \notag
    &\left.\quad=\arctan\!\bigg(\frac{[k_{\perp}^{2}/(2k_{0})]^{2}+\mathrm{Re}[\Omega_{\mathrm{B}}(\mathbf{k}_{\perp},0)]\mathrm{Re}[\Omega_{\mathrm{B}}(\mathbf{k}_{\perp},L)]}{k_{\perp}^{2}/(2k_{0})\times\{\mathrm{Re}[\Omega_{\mathrm{B}}(\mathbf{k}_{\perp},0)]+\mathrm{Re}[\Omega_{\mathrm{B}}(\mathbf{k}_{\perp},L)]\}}\right. \\
    \label{NLPhaseAbsorption}
    &\left.\quad\hphantom{=}\times\tan\!\bigg\{\int_{0}^{L}dz\mathrm{Re}[\Omega_{\mathrm{B}}(\mathbf{k}_{\perp},z)]\bigg\}\bigg),\right.
\end{align}
from which we infer the following linear trend of the transverse displacement $\Delta S(\Lambda)$ in the long-wavelength, superfluid regime:
\begin{equation}
    \label{LinearTrendAbsorption}
    \Delta S(\Lambda)\simeq\frac{1}{2\pi}\arctan\!\bigg(2k_{0}\Delta nL\times\frac{2}{\alpha L}\frac{1-e^{-\alpha L/2}}{1+e^{-\alpha L/2}}\bigg)\Lambda.
\end{equation}

\subsection*{Non-locality model}

So far, we have assumed that the non-linear change of refractive index $\Delta n(\mathbf{r}_{\perp})$ at a given position $\mathbf{r}_{\perp}$ in the transverse plane only depends on the laser intensity at this point, $\propto|\mathcal{E}(\mathbf{r}_{\perp})|^{2}$, and not on the intensity nearby. However, such a local dielectric response may not correctly describe hot atomic vapors, in which the ballistic transport of excited atoms on large length scales induces non-locality \cite{Skupin2007}. Indeed, the coherence between the ground and excited states of an atom, from which the medium non-linear response arises, is more likely to be transported away in hot vapors, as the atomic motion is more significant at large temperatures.

Following \cite{Skupin2007}, we can express the non-local non-linear change of refractive index $\Delta n^{\rm nl}(\mathbf{r}_{\perp})$ as follows: 
\begin{equation}
    \Delta n^{\rm nl}(\mathbf{r}_{\perp}) = n_{2} \int d^{2}\mathbf{r}_{\perp}' G_{\rm b} (\mathbf{r}_{\perp}-\mathbf{r}'_{\perp}) |\mathcal{E}(\mathbf{r}'_{\perp})|^{2},
\end{equation}
where $G_{\rm b}$ stands for the steady-state ballistic response function. By using the convolution theorem, we can then easily rewrite the Bogoliubov dispersion relation (\ref{DispersionRelation}) in the non-local case:
\begin{equation}
    \label{NLDispRelation}
    \Omega_{\rm B}^{\mathrm{nl}}(\mathbf{k}_{\perp}) = \sqrt{\frac{k_{\perp}^{2}}{2 k_{0}} \bigg[ \frac{k_{\perp}^{2}}{2 k_{0}} + 2 k_{0} |n_{2}| \rho_{0} \widetilde{G}_{\rm b}(\mathbf{k}_{\perp}) \bigg]},
\end{equation}
where $\widetilde{G}_{\rm b}$ is the Fourier transform of $G_{\rm b}$. By introducing the ballistic transport length scale $\ell_{\rm b} = u\tau$---where  $u = \sqrt{2 k_{\rm B} T/ m}$ is the most probable speed of the atoms in the transverse plane (at the vapor temperature $T$) and $\tau = 2/\gamma$ is the characteristic decoherence time ---and by calling $\mathrm{erfc}$ the complementary error function, $\widetilde{G}_{\rm b}$ can be written in the following way: 
\begin{equation}
    \widetilde{G}_{\rm b}(\mathbf{k}_{\perp}) = \sqrt{\pi} \frac{e^{1/(k_{\perp}\ell_{\rm b})^2}}{k_{\perp} \ell_{\rm b}} \mathrm{erfc}\bigg(\frac{1}{k_{\perp}\ell_{\rm b}}\bigg).
\end{equation}
The solid lines in Fig.~\ref{fig:ShiftVersus}(c) have been obtained by plugging \eqref{NLDispRelation} into \eqref{HPPhaseLossless}. In the experiment, the vapor temperature was 400 K, leading to a non-local ballistic length $\ell_{\rm b}$ of about 8 $\mu$m. As can be seen in Fig.~\ref{fig:ShiftVersus}(c), non-local effects do not significantly affect the shift $\Delta S$ for such a small value of $\ell_{\rm b}$.  






\bibliography{opticaQF.bib}


\end{document}